\documentclass{article}

\usepackage{arxiv}

\AtBeginDocument{%
  \providecommand\BibTeX{{%
    \normalfont B\kern-0.5em{\scshape i\kern-0.25em b}\kern-0.8em\TeX}}}

\usepackage[utf8]{inputenc} 
\usepackage[T1]{fontenc}    
\usepackage{hyperref}
\hypersetup{colorlinks,allcolors=blue}    
\usepackage{url}            
\usepackage{booktabs}       
\usepackage{amsfonts}       
\usepackage{nicefrac}       
\usepackage{microtype}      
\usepackage{lipsum}
\usepackage{graphicx}
\usepackage{cite}
\usepackage{algorithm2e}
\usepackage{algpseudocode}
\graphicspath{ {./images/} }

\def\eg{\textit{e.g.}}
\def\ie{\textit{i.e.}}

\def\etal{\textit{et al.}}

\newcommand*{\Perm}[2]{{}^{#1}\!P_{#2}}%
\newcommand*{\Comb}[2]{{}^{#1}C_{#2}}%

\title{Affective Computational Advertising Based on Perceptual Metrics}

\author{
 Soujanya Narayana \\
  University of Canberra\\
  Australia \\
  \texttt{soujanya.narayana@canberra.edu.au} \\
   \And
 Shweta Jain \\
  Indian Institute of Technology Ropar\\
  India \\
  \texttt{shwetajain@iitrpr.ac.in} \\
  \And
 Harish Katti \\
  Indian Institute of Science, Bengaluru\\
  India \\
  \texttt{harish2006@gmail.com} \\
  \And
  Roland Goecke \\
  University of Canberra \\
  Australia\\
  \texttt{roland.goecke@ieee.org} \\
  \And
  Ramanathan Subramanian \\
  University of Canberra \\
  Australia\\
  \texttt{ram.subramanian@canberra.edu.au}
}

\begin{document}
\maketitle
\begin{abstract}
We present \textbf{ACAD}, an \textbf{a}ffective \textbf{c}omputational \textbf{ad}vertising framework expressly derived from perceptual metrics. Different from advertising methods which either ignore the emotional nature of (most) programs and ads, or are based on axiomatic rules, the ACAD formulation incorporates findings from a user study examining the effect of within-program ad placements on ad perception. A linear program formulation seeking to achieve (a) \emph{{genuine}} ad assessments and (b) \emph{maximal} ad recall is then proposed. Effectiveness of the ACAD framework is confirmed via a validational user study, where ACAD-induced ad placements are found to be optimal with respect to objectives (a) and (b) against competing approaches.   
\end{abstract}

\keywords{Computational Advertising, Contextual Placement, Linear Programming, Ad Evaluation, Recall
}

\begin{figure}
  \centering
\includegraphics[width=0.99\linewidth]{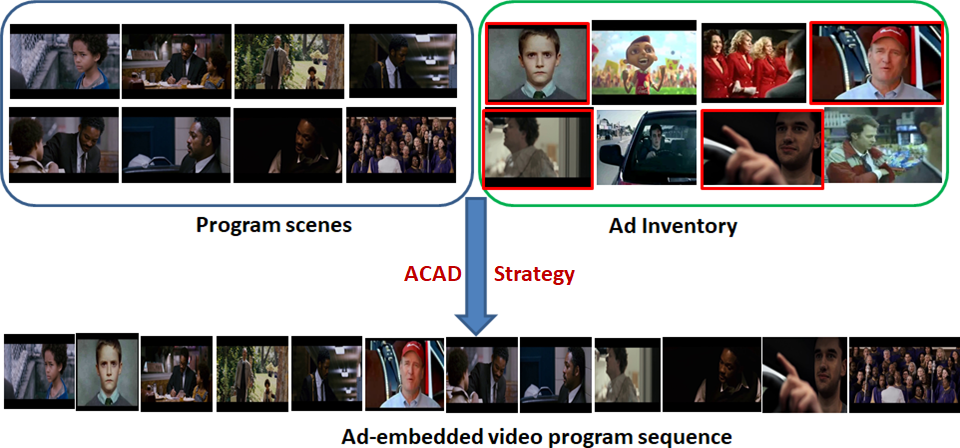}\vspace{-2mm}
\caption{\textbf{ACAD Overview:} Given an emotional \emph{program video} comprising multiple scenes (blue oval), and an \emph{ad inventory} (green oval), ACAD employs linear integer programming to select and embed the requisite number of ads at optimal temporal locations to synthesize a \emph{video program sequence}. ACAD is derived from a user-study examining placement-induced evaluation of affective ads. Ad selections (in red) and placements are determined as a function of (a) ad valence ratings, and (b) the degree of emotional dissimilarity-plus-content similarity between ads and program scenes. ACAD is evaluated via a validational user study.\label{Fig:moneyshot}}\vspace{-.1cm}
\end{figure}
\maketitle

\section{Introduction}
Video advertising has grown by leaps and bounds thanks to multimedia publishers like \emph{Vimeo} and \emph{YouTube}, and popular movie streaming services such as \emph{Netflix} and \emph{Hulu}. As 
humans are emotional, video advertisements ({ads}) are designed to evoke specific emotions in viewers employing a variety of narrative tools to effectively convey a message. Ad designers intend that these emotions become associated with the advertised product (message), and that they influence subsequent consumer behavior\footnote{\emph{Emotional ads work best}: \url{https://www.neurosciencemarketing.com/blog/articles/emotional-ads-work-best.htm}}. 

Ads (commercials) are typically interspersed within longer, often-emotional programs to occasionally distract viewers as well as generate revenue for content publishers. Advertising in broadcast television involves presentation of relevant ads at fixed temporal locations in the program, but a manual mix-and-match strategy would not scale to online content where millions of program videos and ads exist. Algorithmic video-in-video advertising (VIVA) formulations~\cite{cavva,Mei09} attempt to jointly select appropriate ads from an inventory, plus their insertion locations within a program. VIVA as a computational problem needs to balance the conflicting goals of a) minimal \emph{viewing experience} disruption, and b) maximal \emph{ad 
impact}, so that viewers readily recall ads and associated messages over the long term.

CAVVA or computational affective advertising~\cite{cavva} works by matching program scenes and ads with respect to \emph{valence} (positive or negative emotional feeling) and \emph{arousal} (extent of physiological activation)~\cite{Russell1980}, so that viewing experience and ad recall are maximized. CAVVA rules model two axioms, namely, (1) the \emph{emotional inertia} hypothesis~\cite{Broach1995TelevisionPA}, which argues that program emotion lingers and influences evaluation of subsequent ads, and (2) the general tendency of humans to maintain a positive mood. These assumptions preclude CAVVA from inserting ads which strongly contrast with the program valence; for example, a horror movie trailer will rarely be inserted within a sitcom program via the CAVVA framework (see Section~\ref{Sec:Insert_strat}).



Different from~\cite{cavva} which is based on  advertising theories, we expressly examine affective ad evaluation from a user study involving strongly positive and negative programs and ads. The effects of inserting similar/opposite-valence ads within emotional programs are studied to formulate ad-insertion rules, which are realized via linear integer programming to derive the \textbf{a}ffective \textbf{c}omputational \textbf{ad}vertising (ACAD) strategy. ACAD effectiveness is validated via a second user study. Figure~\ref{Fig:moneyshot} summarizes our work, which makes the following research contributions:

\begin{itemize}
\item In contrast to postulate-based advertising strategies~\cite{cavva,Mei09}, the ACAD formulation is derived from a user-study examining the influence of affective ad placements within a positive/negative program. This user study reveals that (a) negative ads are \textit{better recalled} and receive \textit{genuine} ratings when embedded at the tail of the program video, and (b) ads incongruent to the program mood are better recalled overall.

\item Integer linear programming is employed to select the optimal inventory ads and their insertion locations based on the user-study findings. This formulation embeds ads within programs considering (1) valence ratings of the program scenes and ads, and (2) semantic scene--ad similarities. Therefore, low-level (semantic) and high-level (emotional) scene-ad relevance are modeled in the embedding process.

\item A validational user study is performed to evaluate ACAD against competing methods. This study reveals (1) ACAD efficacy towards eliciting genuine ad evaluations and better ad recall, and (2) a trade-off between \emph{ad recall} and \emph{viewing experience}, which should serve as a design guide for future affective advertising works.

\end{itemize}

\section{Related Work}
To position the novelty of our work with respect to the literature, we review related work on (a) affective ad analysis, (b) the influence of program mood on affective ad evaluation, and (c) computational advertising. 

\subsection{Ads as Affective Stimuli} 
Ads are inherently emotional, and emotions play a critical role in effectively conveying a message to viewers regarding lifestyle choices. Emotions can mediate consumer attitudes towards advertised brands~\cite{Holbrook1984,Holbrook1987,Pham2013}, and an advertising strategy that achieves favorable brand (message) evaluation can benefit content publishers and advertisers alike~\cite{cavva}. Adopting the circumplex model~\cite{Russell1980} where emotions are characterized via the \emph{valence} and \emph{arousal} attributes, an affective ad dataset is proposed in~\cite{Shukla2020} and ad affect predicted by examining the content and users. An eye tracking study~\cite{Shukla18} concludes that ad emotions are primarily conveyed via a few {central} scene characters and the visual scene design. The positive impact of emotional cues towards ad and movie scene recall is confirmed in~\cite{FRIESTAD19931} and~\cite{Subramanian14}.    


\subsection{Program Influence on Ad Perception} 
As ads are interspersed within programs, the impact of program mood on ad evaluation has been studied extensively. Program \emph{context}, characterized by audio-visual and emotional program-ad (dis)similarity, impacts ad evaluation. Kamins~\etal~\cite{Kamins91} focused on how commercial {liking}, and consequently, user purchase intent is affected by program mood; they observed the \emph{emotional inertia} effect, \ie, happy ads are perceived favorably within a happy program while sad ads are better liked in a sad context. Broach~\etal~\cite{Broach1995TelevisionPA} noted that both program arousal (asl) and valence (val) impact ad perception. Specifically, positive ads are evaluated favorably within a highly arousing positive program or a mildly arousing negative program. These findings strengthen the emotional inertia theory~\cite{Meyers1993} as lingering emotion under high asl adversely impacts focus on opposite val ads; greater cognitive resources available under low asl result in ads being viewed as more positive/negative within a sad/happy program. Only positive ads are nevertheless studied in~\cite{Broach1995TelevisionPA}. 

Varied findings exist regarding user evaluation of ads (in)congruent to the program mood. A study involving only two programs and two ads~\cite{Aylesworth98} concludes that it may be optimal to show promotional ads within positive programs, and messaging campaigns within negative programs. Other studies nevertheless point to increased recall of incongruent ads; \eg,~\cite{Feltham1994} notes that when one watches an arousing program, emotionally inconsistent ads tend to stimulate internal processing and are therefore better recalled. Similarly,~\cite{Furnham13} notes that health and safety-related messages are better recalled within a sitcom program. The positive influence of program arousal on ad recall, where users are free to watch or skip ads, is noted in~\cite{Moorman07}. Shapiro~\etal~\cite{shapiro2013understanding} argue that program arousal and valence independently influence ad recall. Asl impacts information processing capacity (deeper processing under low asl, shallow processing under high asl), while val influences the nature of processing (logic-based but superficial under high val, and attention to detail under low val). As information processing is key to brand evaluation and attitude, this study has design implications for computational VIVA strategies. 

\subsection{Strategic VIVA}
To exploit the surge in Internet video and benefit both content publishers and advertisers, computational advertising strategies have been proposed.
An appraisal of the computational advertising field~\cite{Yun2020} emphasizes the need to account for consumer behaviors during advertising design, execution and evaluation. However, VIVA approaches typically examine user behaviors only at the evaluation stage.  VideoSense~\cite{Mei09} inserts ads at temporal points of low involvement (\eg, shot discontinuities), and presents ads which are audio-visually similar to the preceding video scene. The twin objectives of optimal ad selection and placement are modeled via non-linear integer programming (NLIP). While subjective evaluations are presented to validate the approach, emotional program-ad relevance is ignored in VideoSense. 

The CAVVA framework~\cite{cavva} improves over VideoSense by modeling emotional relevance for program-ad matching. Matching rules are derived from~\cite{Broach1995TelevisionPA}, and by postulating that the embedded ads should (a) minimally perturb user emotional state and help attain/maintain a positive mood, and (b) maximally engage viewers. Consistently, low-val ads are inserted after negative-val scenes, while high-val ads are embedded between positive-val scenes. Ad selections and placements are achieved via NLIP, and CAVVA achieves superior viewing experience and ad recall than VideoSense. Rules modeling emotional inertia preclude CAVVA from embedding ads incongruent to program mood. 

\subsection{Inference Summary \& Research Questions}
We note from the literature that (a) while a few works have measured ad-induced user behaviors, user assessments have hardly been incorporated for VIVA design; (b) even as marketing studies have examined the effects of program mood on affective ad evaluation, very few VIVA works embed ads based on program-ad emotional relevance, and (c) VIVA methods either focus on content-based or mood-sustaining advertising. Embedding ads strongly incongruent to the program mood is therefore not possible.

To address these issues, we expressly perform an exploratory user study involving both positive and negative programs and ads to understand (a) the impact of program mood on ad evaluation and recall, and (b) whether certain temporal placements are desirable for positive and negative val ads. Findings convey that optimal commercial evaluation and recall are achieved when (1) low val ads are placed towards the program {tail}, and (2) inserted ads are emotion-wise dissimilar but content-wise similar to the (preceding) scenes. The study employed to derive ACAD rules is described below.

\section{Materials and Methods}\label{Sec:MandM}

To examine the influence of program mood on ad evaluations, we performed a user-study with two programs (one happy and another sad) embedded with both positive and negative emotional ads as described below.  


\subsection{Materials}\label{Exp:Mat}
  \textbf{Programs:} For this work, we utilized the affective ad dataset and two program videos (PVs) presented in~\cite{Shukla2020}. The first PV is a sequence from the movie {\emph{In Pursuit of Happiness}} ({IPOH} of {drama} genre), while the other PV is episode 1 from the {\emph{Friends}} Season 2 sitcom. Both PVs comprise eight scenes and are of roughly similar lengths (mean scene length of 110$\pm$44s for {IPOH} vs 119$\pm$69s for Friends). These two PVs were considered as they depicted self-contained episodes, and were fairly representative of online videos\footnote{Mean \textit{Youtube} video length = 11.7 minutes (as of Dec 2018, Statista.com).}. One can infer from~\cite{Shukla2020} that both PVs are sufficiently arousing. While the \emph{Friends} PV is happy mood-inducing, the {IPOH} PV detailing the travails of a father and son is sad mood-eliciting. 
  
\noindent \textbf{Ads:} To examine how emotional ads are perceived within an emotional program, we used 14 ads--7 high asl-high val (HAHV) and 7 high asl-low val (HALV) ads from~\cite{Shukla2020}. The HAHV ads are promotional in nature, while the HALV ads are public messages conveying the ill effects of unhealthy and hazardous lifestyle choices such as overspeeding, smoking, drug and physical abuse.

\noindent \textbf{Video program sequences:} As both programs and the selected ads are deemed to be sufficiently arousing, we exclusively examined the effect of program and ad valence on commercial evaluation. For the user study, four (2 HAHV + 2 HALV) of the 14 ads were randomly chosen and inserted within each PV. Consistent with the VideoSense~\cite{Mei09} and CAVVA~\cite{cavva} methods, we embedded ads uniformly across each PV. As both PVs had seven scene-change points where ads could be inserted, chosen ads were placed at change-point 1, 3, 5 and 7. Cumulatively, we examined four different ad placement configurations depicted in Fig.~\ref{fig:Illustration}, where HV (LV) ads are placed at the \emph{head} (\emph{tail}), \emph{tail} (\emph{head}), \emph{ends} (\emph{middle}) and \emph{middle} (\emph{ends}) of the PV. These four configurations summarize \emph{{meaningful}} placements of the 2 HV + 2 LV ads given four insertion locations (totally $\Perm{4}{4}$ possibilities). The ads were 0.5--1 minute long ($\mu = 45$s, $\sigma = 14.5$s). The ad-embedded programs or \emph{video program sequences} (VPSs) were $\approx 18$ minutes long ($\mu = 18.3, \sigma = 0.6$ minutes). 

\subsection{Participants}
24 university students (15 male) aged between 18--39 years ($\mu = 23.4, \sigma=4.9$) conversant with English language programs participated in the study. This demographic accounts for 21\% of the global audience accessing online video websites\footnote{as per  hootsuite.com}, whom advertising approaches should most effectively target. All users were provided with a definition of stimulus-related terms such as PV and VPS, and the {valence} and {arousal} emotional attributes. Participants provided written informed consent as per ethics approval, and were paid a token fee for their time and effort. 


\subsection{Procedure}
The IPOH and Friends VPSs were presented on a computer screen. Upon viewing, users rated each ad on a 1--5 scale for val (\emph{very unpleasant}--  \emph{very pleasant}) and asl (\emph{boring} -- \emph{exciting}) via a radio button. On VPS completion, users rated the program video for asl and val, and the ad-embedded VPS for viewing experience (VX) on a 1--5 scale (\emph{very bad} -- \emph{very good}). To test visual ad recall, viewers were shown thumbnails of the 14 inventory ads, and asked to identify the four {viewed} ads in any order. This protocol resulted in $24 \times 2 \times 4 = 192$ user asl, val ratings being available for ads, and $24 \times 2 = 48$ scores being available for program asl and val, ad recall, ad asl and val per placement condition and the VPS VX. Including preparation time and a break between the two VPSs, $\approx$50 minutes were required for the experiment. We will publish the user-study data and the ACAD implementation for research purposes upon paper acceptance. 

\section{User Data Analyses}\label{Sec:US_anal}
In this section, we examine (i) asl,val ratings for PVs, (ii) within-program ad evaluations in terms of asl, val and recall scores, focusing on both general and placement-specific trends, and (iii) ad-placement effects on VPS viewing experience.    

\begin{figure}[!htbp]
\centering
\includegraphics[width=0.8\linewidth, height=2.8cm]{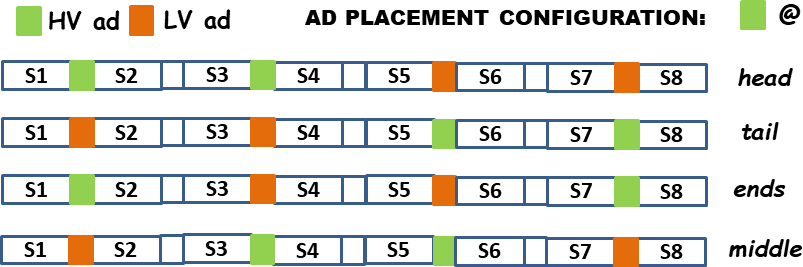}\vspace{-.2cm}
\caption{Ad placement strategies examined for our user study. S1--S8 denote the eight PV scenes.}\label{fig:Illustration}\vspace{.1cm}
\centering
\includegraphics[width=0.4\linewidth, height = 3.5cm]{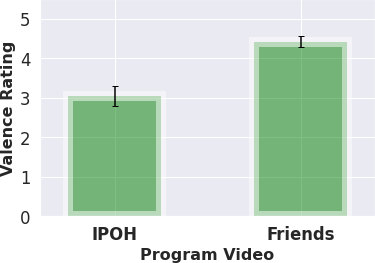}\hspace{5mm}\includegraphics[width=0.4\linewidth, height = 3.5cm]{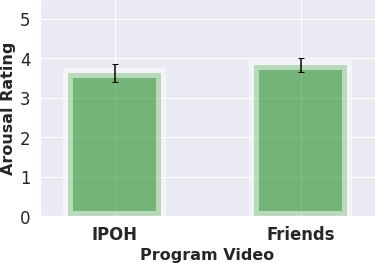}\vspace{-.2cm}
\caption{(L) Mean val and (R) mean asl ratings obtained for the two PVs. Error bars denote unit standard error of mean.}\label{fig:PV_VArat} 
\end{figure}

\subsection{Asl,val Ratings for PVs}
Fig. \ref{fig:PV_VArat} presents the mean val (left) and asl (right) user ratings for the two PVs. The IPOH and Friends PVs respectively elicited  mean val scores of 3 and 4.4, which were significantly different as per an unequal variance $t$-test $(t(46) = 4.827, p<0.000001)$. Both PVs evoked similar asl ratings-- 3.6 for IPOH vs 3.8 for Friends $(t(46) = 0.714, \mbox{n.s.})$. These observations are similar to \cite{Shukla2020}, and the Friends and IPOH programs are termed as the happy and sad PVs respectively. 


\begin{figure}[!htbp]
\centerline{\includegraphics[width=0.45\linewidth]{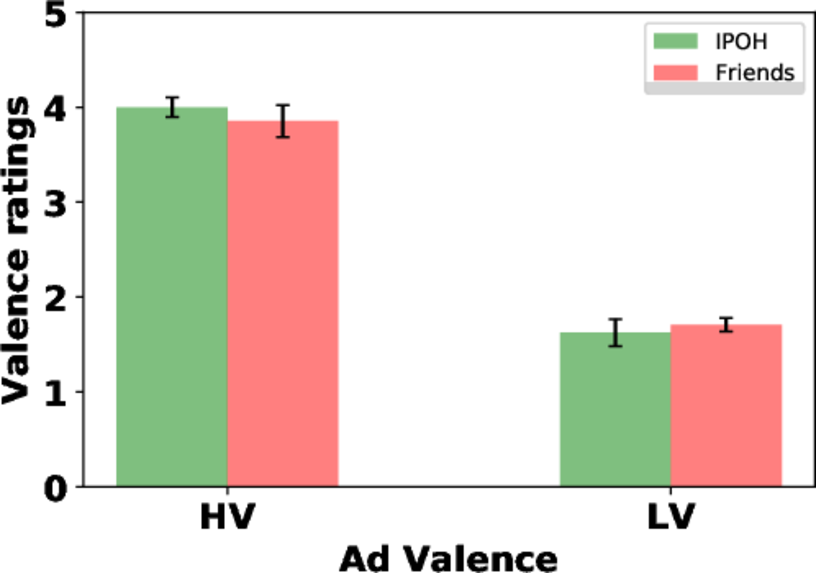}\hspace{.1cm}\includegraphics[width=0.45\linewidth]{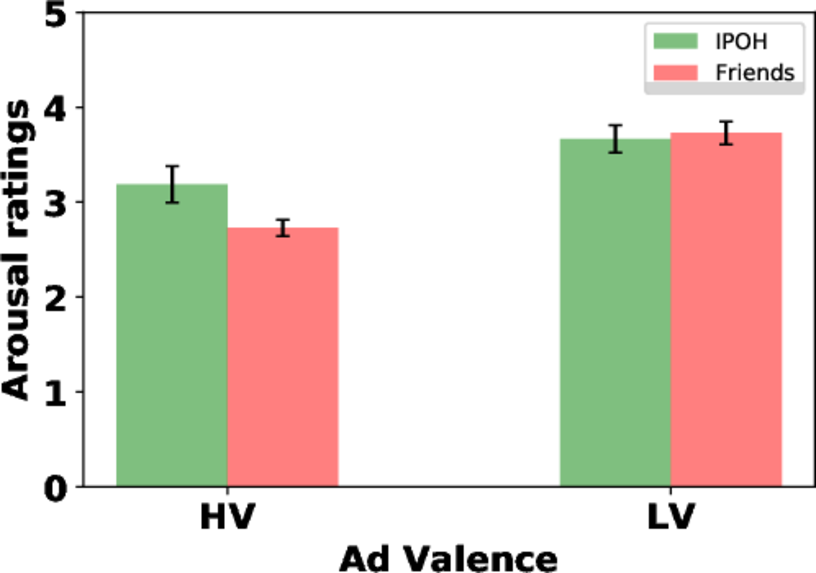}}
\vspace{-.2cm}
\caption{Mean ad val,asl ratings when shown within the two PVs. Error bars denote unit standard error of mean.}\label{fig:AdVAE_ratings}
\vspace{2mm}
\centerline{\includegraphics[width=6cm,height=3cm]{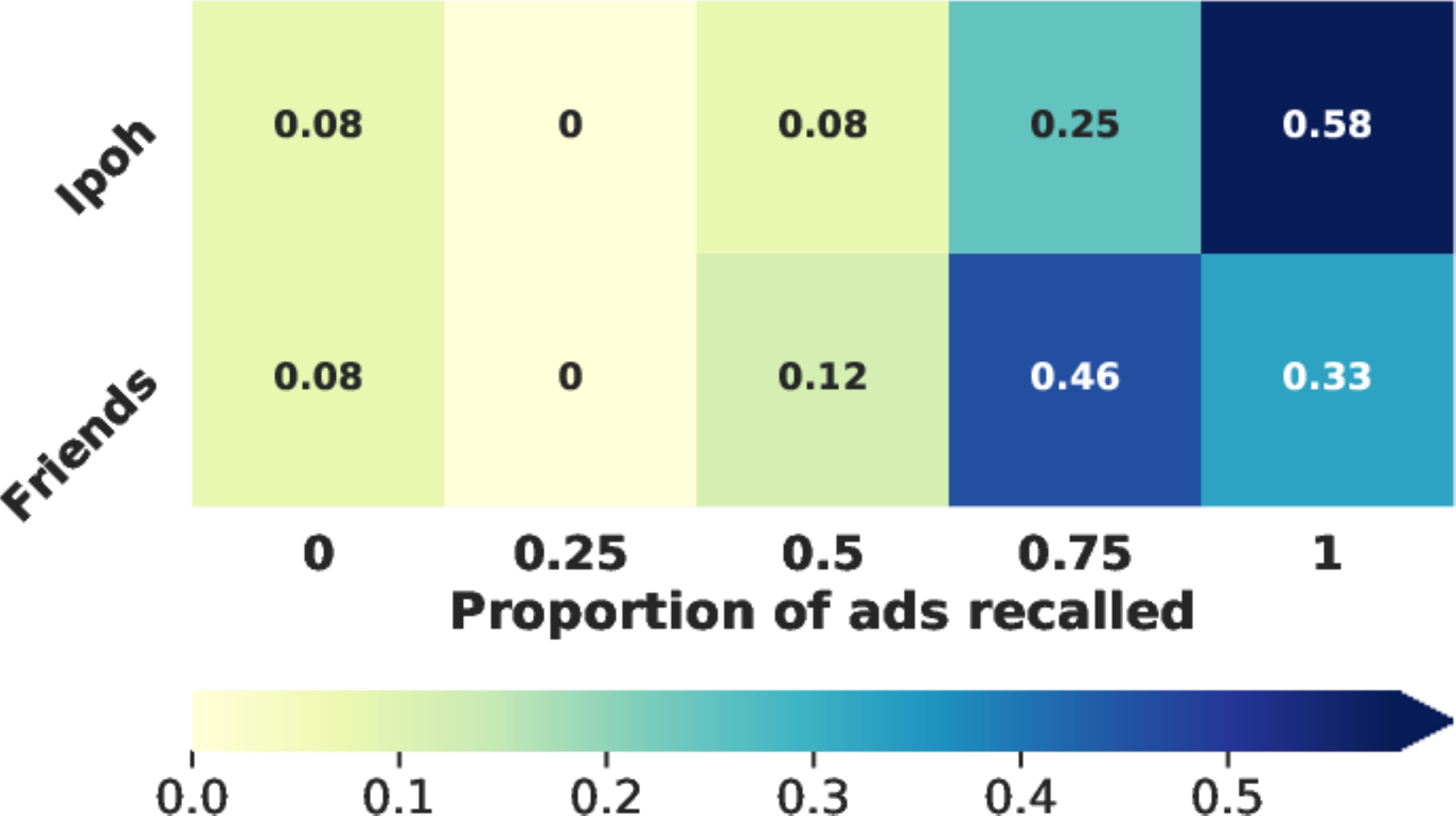}\hspace{0.5cm}\includegraphics[width=6cm,height=3cm]{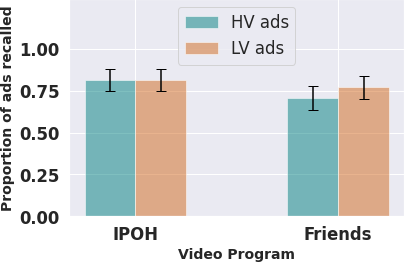}}\vspace{-.2cm}
 \caption{Ad recall: (left) Heatmap showing fraction of users for varying ad recall rates. (right) Program-wise recall rates for high and low-valence ads. Error bars denote standard error of mean. Viewed best in color and under zoom.}
    \label{fig:Recall_res}\vspace{-.5cm}
\end{figure}

\subsection{Impact of PV Mood on Ad asl,val Ratings}
To understand how program-induced mood impacts ad evaluations, we examined user asl,val ratings for HV and LV ads inserted within the happy and sad PVs. Fig.~\ref{fig:AdVAE_ratings} presents mean asl,val scores received by the HV and LV ads within the positive and negative programs. With respect to {valence} (Fig.~\ref{fig:AdVAE_ratings}~(left)), HV ads expectedly elicited higher val ratings than LV ads within both programs. HV ads elicited mean val scores of 4$\pm$0.2 and 3.9$\pm$0.3 when shown within the sad and happy PVs, indicating that HV ads were perceived as \emph{more pleasant} in a sad PV. Conversely, LV ads evoked a mean val score of 1.6$\pm$0.3 and 1.7$\pm$0.1 within the IPOH and Friends PVs, implying that LV ads were seen as \emph{less unpleasant} in a happy program. However, perceived differences in ad val for the happy and sad programs were insignificant as per a one-way analysis of variance (ANOVA) test $(F(1,188) = 0.06, \mbox{n.s})$.    

In terms of {arousal} impressions (Fig.~\ref{fig:AdVAE_ratings}~(right)), LV ads evoked higher asl vis-\`a-vis HV ads in general. HV ads shown in the IPOH and Friends PVs received asl ratings of 3.2$\pm$0.3 and 2.7$\pm$0.2 respectively, while corresponding LV ads achieved higher asl scores of 3.7$\pm$0.3 and 3.8$\pm$0.2. A two-way ANOVA confirmed the main effect of ad val $(F(1,188) = 18.77, p<0.00001)$, but the negligible impact of program val $(F(1,188) = 1.34, \mbox{n.s})$~on asl scores. Post-hoc $t$-tests however revealed slightly different asl scores for HV ads within the happy and sad PVs $(t(94) = -1.75, p=0.083)$. 

\subsection{Ad Recall Analysis}
Visual ad recall critically influences brand attitude~\cite{cavva,FRIESTAD19931}. Fig.~\ref{fig:Recall_res}~(left) presents ad recall for the happy and sad PVs irrespective of ad valence. The heatmap depicts the proportion of users achieving a given ad recall rate for the happy and sad programs. Evidently, ad recall rates were higher for the sad program. 58\% users recalled all ads within the sad program, while only 33\% users remembered all ads in the happy program. Fig.~\ref{fig:Recall_res}~(right) presents program-wise recall of the HV and LV ads. As in Fig.~\ref{fig:Recall_res}~(left), superior recall of both HV and LV ads is achieved for the sad PV (mean recall rate or MRR of 0.81 for both HV and LV ads), as compared to the happy PV (MRR of 0.77 for LV vs 0.71 for HV ads). ANOVA tests to examine the effects of {PV} and {ad valence} on ad recall did not reveal any effects.


\subsubsection{Discussion}
Observations from the user study are largely consistent with prior findings. With respect to ad asl,val ratings, LV ads are more arousing in general. Negative ads elicited higher asl scores as per~\cite{Shukla2020}, and this holds even within a program context as seen in Fig.~\ref{fig:AdVAE_ratings}(right). In terms of valence, HV ads elicited higher scores when shown within a sad program, while LV ads evoked higher val scores within a happy program. That ads are better remembered in the negative program context is also known; prior studies~\cite{Kensinger,Rimmele11} note that negative mood induces better recall, and greater attention to detail~\cite{shapiro2013understanding}. 
Greater memorability for negative stimuli~\cite{Kensinger,Rimmele11} explains why LV ads are better recalled irrespective of program mood. That HV ads are better recalled in the sad PV context can be attributed to the {contrastive} effect noted in prior studies~\cite{Feltham1994,Furnham13}. While these trends suggest that a promotional ad may have more impact when shown in an arousing sad program, we set out to examine if certain placements are more optimal for HV and LV ads given the program mood.   

\begin{figure}[!t]
\centering
    \includegraphics[width=0.35\linewidth]{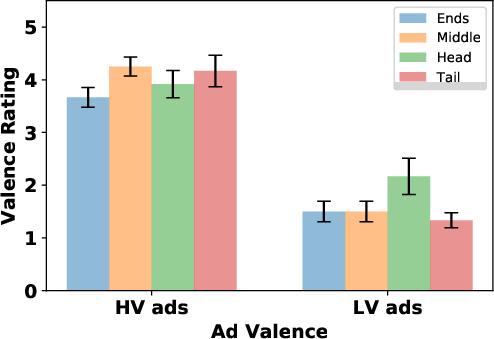}\hspace{0.4cm}\includegraphics[width=0.35\linewidth]{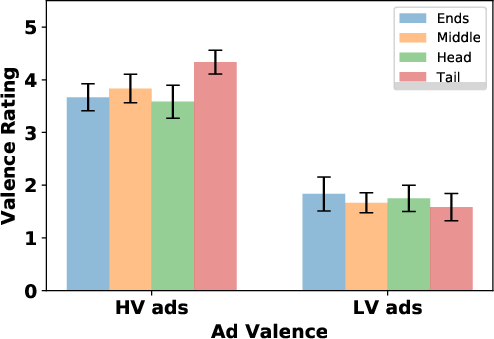}
        \vspace{-2mm}\caption{Ad val scores by position in sad (L) and happy (R) PVs.}\label{fig:AdPos_Val}
    \vspace{2mm}
        \centering
        \includegraphics[width=0.35\linewidth]{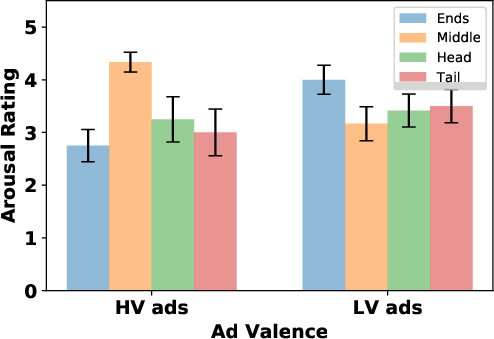}\hspace{0.4cm}\includegraphics[width=0.35\linewidth]{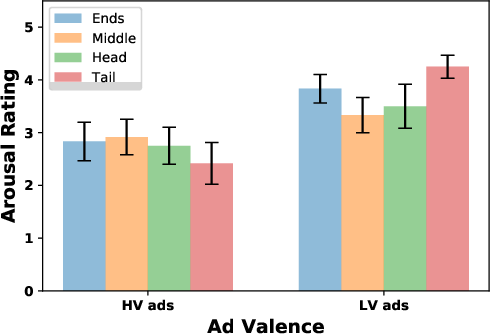}\vspace{-.2cm}
        \vspace{-1mm}\caption{Ad asl scores by position in sad (L) and happy (R) PVs.}\label{fig:AdPos_Asl}\vspace{-4mm}    
\end{figure}

\subsection{Ad Placement Effects}
\begin{sloppypar}
As the primary objective of this work is to examine whether some within-program ad placements are \emph{better} than others for ads (in)congruent to the program mood, and devise a computational advertising strategy therefrom, we examined if ad evaluations differed for the varied placements shown in Fig.~\ref{fig:Illustration}.
\end{sloppypar}

\subsubsection{Asl,val impressions:} Ad positioning induced marginally different user emotional ratings as detailed below. 

\noindent \textbf{Valence:} Fig.~\ref{fig:AdPos_Val} presents val scores elicited for varied ad placements within the two PVs (error bars denote standard error of mean). Within the sad PV (Fig.~\ref{fig:AdPos_Val} (left)), a two-way ANOVA on val ratings with \emph{ad valence} and \emph{ad position} as factors failed to reveal any ad positioning effect $(F(3,88) = 1.37, \mbox{n.s})$, but revealed a slight ad valence--position interaction effect $(F(3,88) = 2.39, p=0.074)$. HV ads induced better val impressions at the \emph{tail} and \emph{middle} (mean val = 4.2 for both), than at the \emph{head} (mean val = 3.9) or \emph{ends} (mean val = 3.7).  
LV ads were deemed most unpleasant at the \emph{tail} (mean val = 1.3), and least unpleasant at the \emph{head} (mean val = 2.2), and this difference was marginally significant ($t(46) = -1.933, p=0.06$); no differences were noted between the \emph{ends} or \emph{middle} placements (mean val = 1.5). 
Within the happy PV (Fig.~\ref{fig:AdPos_Val}~(right)), a two-way ANOVA again failed to reveal any effect of ad positioning, but revealed a significant ad valence--position interaction effect $(F(3,184) = 3.3107, p<0.05)$. HV ads were deemed more +ve at the \emph{tail} (mean val = 4.3) than at the \emph{middle} (mean val = 3.8), \emph{ends} (mean val = 3.7) or \emph{head} (mean val = 3.6). Slightly different val distributions were noted for the \emph{tail} and \emph{head} positions ($t(46) = 1.947, p=0.066$), and also for the \emph{tail} and \emph{ends} ($t(46) = 1.956, p=0.063$). LV ads were perceived as most -ve when placed at the \emph{tail} (mean val = 1.6), and least -ve at the \emph{ends} (mean val = 1.8), with the \emph{head} (mean val = 1.75) and \emph{middle} (mean val = 1.7) positions achieving moderately negative evaluations. val-score differences  among these positions were however insignificant.  

\noindent \textbf{Arousal:} Fig.~\ref{fig:AdPos_Asl} presents asl ratings elicited for varied ad placements within the negative and positive PVs, with error bars denoting standard error of mean. Within the sad PV (Fig.~\ref{fig:AdPos_Asl}~(left)), a two-way ANOVA revealed a significant ad valence--position interaction effect $(F(3,184) = 4.614, p<0.005)$. HV ads evoked most asl in the \emph{middle} (mean asl = 4.3), and least asl at the \emph{ends} (mean asl = 2.8). \textit{Middle} ad ratings differed significantly with respect to the \textit{ends} ($t(46) = 4.423, p<0.0005$),  \emph{head} ($t(46) = 2.315, p<0.05$) and \emph{tail} ($t(46) = 2.766, p<0.05$). LV ads evoked most asl at the \emph{ends} (mean asl =4.0), and least asl in the \emph{middle} of the PV (mean asl =3.2). Within the happy PV (Fig.~\ref{fig:AdPos_Asl}~(right)), ANOVA revealed an ad valence--position interaction effect $F(3,184) = 4.44, p<0.005)$. HV ads in the \emph{middle} (mean asl = 2.9) and \emph{tail} (mean asl = 2.4) elicited distinct asl ratings  ($t(46) = -2.943, p<0.01$). LV ads evoked most and least asl at the \emph{tail} (mean asl = 4.3) and \emph{middle} (mean asl = 3.3), and these scores differed significantly ($t(46) = 8.983, p<0.000001$). 


\begin{figure*}[!t]
\centering
\begin{minipage}{.56\textwidth}
   \centering
    \includegraphics[width=0.495\linewidth,height=2.5cm]{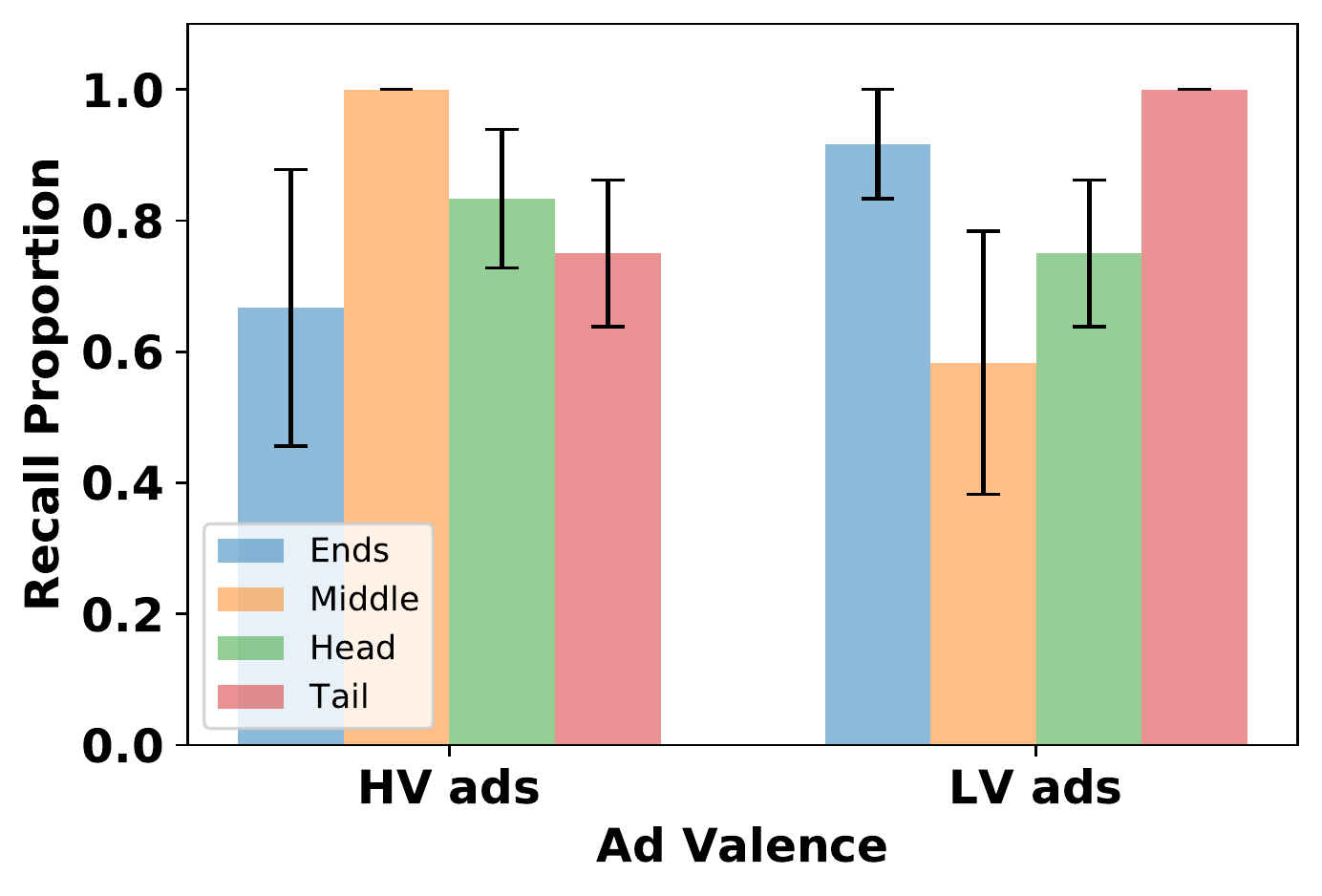}\hspace{0.2mm}\includegraphics[width=0.495\linewidth,height=2.5cm]{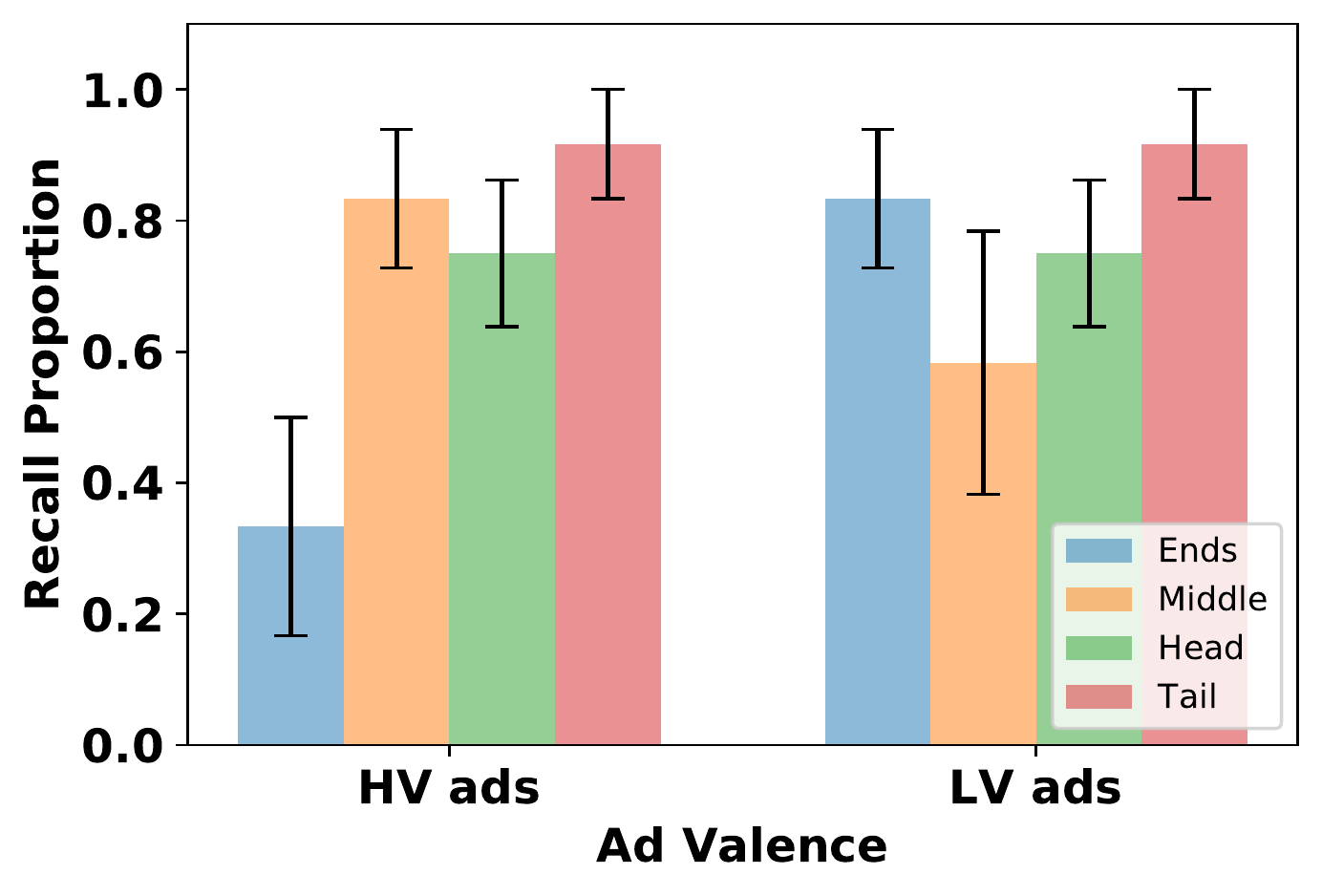}
        \vspace{-7mm}\caption{Ad recall by position in sad (L) and happy (R) PVs. Error bars denote standard error of mean.}\label{fig:AdPos_Rec}
\end{minipage}\hspace{2mm}
\begin{minipage}{.42\textwidth}
 \centering
\includegraphics[width=0.7\linewidth,height=2.5cm]{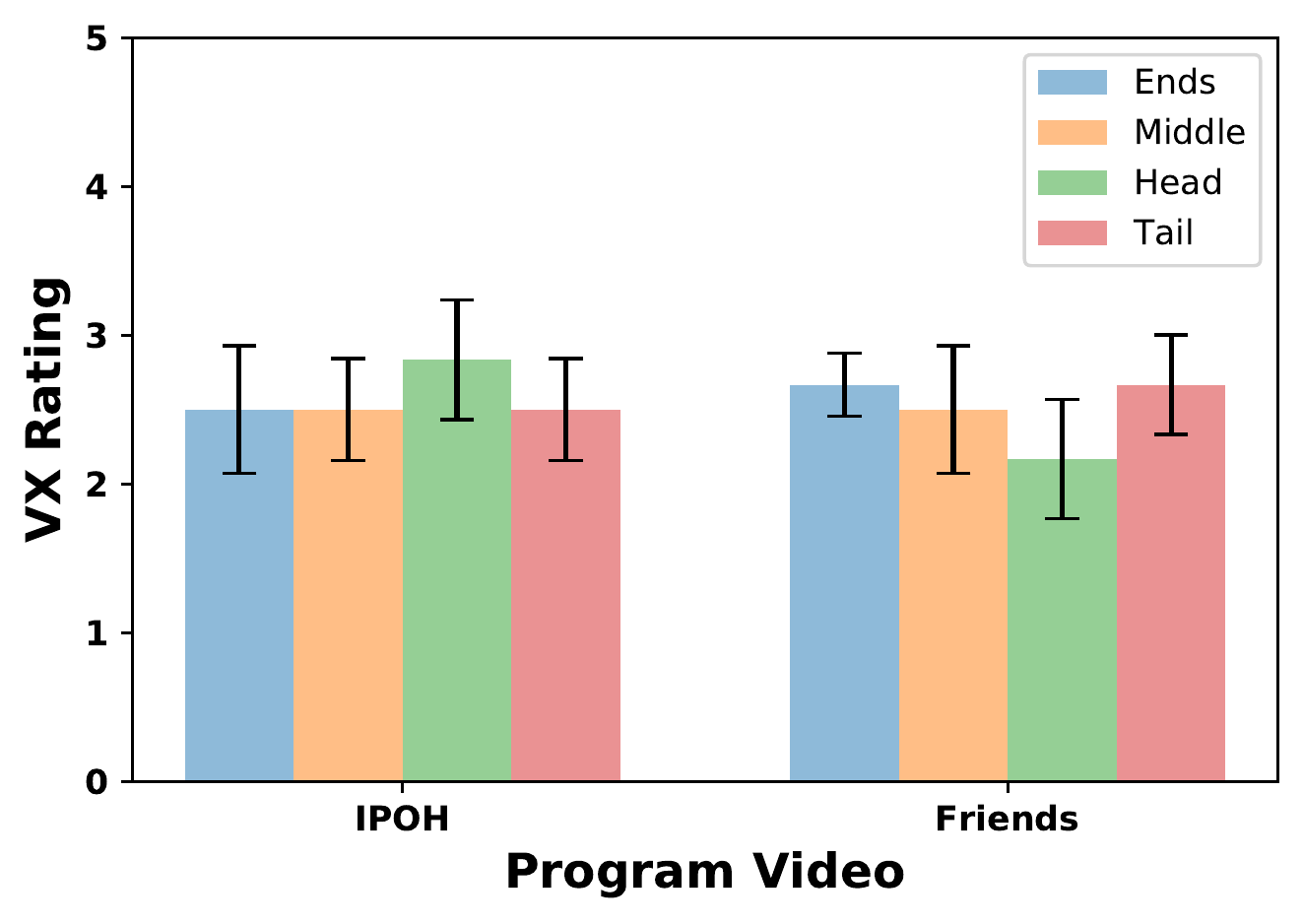}\vspace{-4mm}
        \caption{VX for HV ad placements within the two PVs. Error bars denote standard error of mean.}\label{fig:AdPos_Exp}
\end{minipage}
\vspace{-3.5mm}   
\end{figure*}

\subsubsection{Ad Recall} For both the sad (Fig.~\ref{fig:AdPos_Rec} (left)) and happy (Fig.~\ref{fig:AdPos_Rec} (right)) programs, we noted a marginal impact of ad positioning on recall. For the sad PV, ANOVA revealed a significant ad valence--position interaction effect on visual recall ($F(3,40) = 3.18, p<0.05)$. Highest recall of HV ads was noted in the VPS \emph{middle} (MRR = 1), and lowest at the \emph{ends} (MRR = 0.67). For LV ads, recall was higher at the \emph{tail} (MRR = 1) than the \emph{middle} (MRR = 0.58) and \emph{head} (MRR = 0.75) positions. LV ad recall rates differed marginally for the \emph{tail} vs \emph{middle} ($t(46) = 2.076, p=0.093$), and for \emph{tail} vs \emph{head} ($t(46) = 2.236, p=0.076$). For the happy PV, ANOVA revealed a marginal effect of ad positioning on recall rate $(F(3,40) = 2.35, p=0.087)$, but a significant interaction effect $(F(3,40) = 3.06, p<0.05)$. HV ads were recalled best at the \emph{tail} (MRR = 0.92), and worst at the \emph{ends} (MRR = 0.33), and these recall rates differed significantly ($t(46) = -3.13, p<0.05$). LV ads were also recalled best at the \emph{tail} (MRR = 0.92) and worst in the \emph{middle} (MRR = 0.58), but the differences were marginal.
 
\subsubsection{Viewing experience (VX)}
As the CAVVA~\cite{cavva} framework is designed to perform ad insertions that sustain user mood to ensure a pleasant viewing experience, we examined if VX scores differed with varying ad-placements. Fig.~\ref{fig:AdPos_Exp} summarizes the impact of HV ad-placements within the sad and happy programs (see Fig.~\ref{fig:Illustration} for inferring corresponding LV ad placements). For the sad PV, superior VX (mean VX = 2.8) was reported when HV (LV) ads were placed at the VPS \emph{head} (\emph{tail}), with the other three configurations eliciting identical mean VX scores of 2.5. Within the happy PV, very similar VX scores were reported for different configurations; overall, VX score differences were insignficant.

\subsection{Inferences \& ACAD Design Rules} 
Our analysis reveals that within program ad-placements have some effect on affective impressions and recall rates, with ANOVA tests largely revealing a marginal effect of ad-positioning, or an interaction effect between ad valence and positioning. Interestingly, trends common to both HV and LV ads are apparent irrespective of program mood such as: (a) HV ads being perceived as most pleasant when placed at the VPS \emph{tail} or \emph{middle}, and LV ads perceived as most unpleasant when placed at the VPS \emph{tail}; (b) HV ads being recalled best at the VPS \emph{middle} and \emph{tail}, and LV ads achieving best recall at the VPS \emph{tail}, and (c) HV ads evoking maximal asl when placed in the \emph{middle} of the VPS. 

Overall, our study confirms that both \emph{program mood} and \emph{ad placement} impact ad evaluation and recall. Ad placement effects convey that it may be beneficial to present (especially LV) ads later in the VPS, as the \emph{middle} and \emph{tail} embeddings generally achieve genuine val ratings than at the \emph{ends} or the \emph{head}. As our study involves presentation of ads strongly (dis)similar to the program mood, our findings suggest that users are able to evaluate ads more genuinely upon sufficient program exposure. Programs are likely to maximally attract viewers at the beginning, as they begin to engage with the content; ads presented under such high asl conditions are not adequately processed by users as per findings in~\cite{Broach1995TelevisionPA}. With sufficient program exposure, users tend to transit to a moderately aroused state, where both positive and negative messages are well received as conveyed by minimal VX variations across placements.

Given that the success of an advertising/awareness campaign hinges on a positive message being viewed most favorably, and conversely, a negative message being viewed most unfavorably by consumers, we hypothesize that a computational VIVA framework should generate ad embeddings that achieve the most \emph{genuine valence} evaluations. Secondly, the ads should also have maximum \emph{recall rates}, so that they influence purchase intent~\cite{cavva}. Consistently, the ACAD strategy should result in (a) genuine ad valence evaluations, and (b) maximal ad recall. This imposes  (1) LV ads to be embedded at the VPS \emph{tail} where they achieve the most genuine assessments and maximum recall; HV ads are also well received in the VPS \emph{middle} where they evoke high asl and achieve good recall. (2) Ads incongruent to the program mood are well recalled (Fig. \ref{fig:Recall_res} (right)); LV ads achieve high recall irrespective of program mood, while HV ads are better recalled in the negative program context. (3) In line with the CAVVA~\cite{cavva} and VideoSense~\cite{Mei09} strategies, we hypothesize that content-based ad-scene relevance will facilitate ad recall (\eg, a kitchen commercial may be better remembered after a dinner scene). ACAD   
realizes rules (1)--(3) via a linear program formulation as elaborated below.

\section{ACAD formulation}
Let $\{S_1,S_2,\ldots,S_N\}$ denote the $N$ program scenes, onto which ads can be inserted at $M$ slots ($M=N-1$ if scene transition points are candidate insertion slots). Let $\{A_1,A_2,\ldots,A_P\}$ denote the $P$ inventory commercials from which $K$ ads are to be chosen, so as to embed an equal number of HV and LV ads in the program. We introduce selection-cum-insertion binary variables $y_{ij} \in \mathbb{R}^{N \times P}$, where $y_{ij}=1$ if ad $A_j$ follows program scene $S_i$ and $y_{ij}=0$ otherwise. Let $Y$ be a $N \times P$ matrix containing these binary variables. As our study involves arousing programs and ads, we only consider val scores in the program scene-ad matching process. Let $val^{S_i}$ and $val^{A_j}$ denote the valence evoked by scene $S_i$ and ad $A_j$, where $val^{S_i},val^{A_j} \in [0,1]$. Further, let $val^S \in \mathbb{R}^N$ and $val^a \in \mathbb{R}^P$ respectively denote the vectors denoting scene and ad valence scores.  Our objective is to then to come up with the matrix $Y$ so as to maximize the reward function $RF(val^S,val^asl, \alpha, \beta)$ with trade-off parameters $\alpha, \beta$, where 
\vspace{-1.5mm}
\small
\begin{equation}
{RF(val^S, val^asl, \alpha, \beta) = \alpha\sum_{i=1}^M i \sum_{j=1}^P y_{ij}(1-val^{A_j}) +  
\beta\sum_{i=1}^M \sum_{j=1}^P y_{ij}(\vert val^{S_i} - val^{A_j} \vert)rel(S_i,A_j)} 
\label{eq:RF}
\end{equation}


\noindent s.t. $\sum_{j=1}^P y_{ij} = 1 \ \forall i;\quad \sum_{i=1}^M \sum_{j=1}^P y_{ij} = K$; 
\quad $\sum_{i=1}^{\lceil M/K \rceil} y_{ij} = 1 \ \forall j$ \ and \quad $\alpha + \beta =1$ \\
\normalsize

Here, $rel(S_i,A_j)$ models the content-wise similarity between scene $S_i$ and ad $A_j$. Decoding Eq.~(\ref{eq:RF}), the $\alpha$ term enforces LV ads, for which $(1-val^{A_j})$ is high, to be embedded later in the VPS. The $\beta$ term encourages emotionally dissimilar (so that $val^{S_i} - val^{A_j}$ is large) but content-wise similar scene-ad combinations. The constraints, in order, specify that (a) only one ad can be filled per ad-slot; (b) total number of embedded ads is $K$ and (c) the inserted ads be uniformly distributed over the VPS as in~\cite{cavva,Mei09}, requiring one ad to be placed over every $[M/K]$ slots, where $\lceil . \rceil$ denotes the \emph{ceil} function operator. Equation (\ref{eq:RF}) represents a binary integer linear program which is solved via Algorithm~\ref{bf_acad}.

\subsection{Brute-Force ACAD Algorithm}
\small
\begin{algorithm}[!ht]
\SetKwInput{KwInput}{Input} 
\SetKwInput{KwOutput}{Output}  
\KwInput{Program scenes valence $\{val^{S_1},val^{S_2},\ldots,val^{S_N}\}$, inventory ads valence $\{val^{A_1},val^{A_2},\ldots,val^{A_P}\}$. Number of ads to insert: \emph{K} and number of ad slots: \emph{M}, Relevence matrix $rel(S_i, A_j)$ denoting the relevance score between scene $i$ and ad $j$, Parameters $\alpha$ and $\beta$.}
\KwOutput{Optimal ad-embedded VPS schedule.}
\begin{enumerate}
    \item Initialize a 1-dimensional array $opt\_allo$, which denotes the optimal ad sequence, with $M$ zeros.
    \item Create a matrix $C$ with $\Comb{P}{K}$ rows and $K$ columns, whose rows consist of all possible subsets of $\{1,\ldots,P\}$ with $K$ elements. Retain only those rows in $C$, which have a balanced sampling of HV and LV ads from the inventory.
    \item For each row $r$ in $C$ which denotes a valid ad sampling, create a matrix $L$ with $M!$ rows and $M$ columns, where each row is a permutation of the ad choices in $r$. We fill these $M$ slots with $K$ ads, such that constraint (b) is satisfied.
    \item For each row $j$ in $L$, compute reward function $RF(val^S,val^asl, \alpha, \beta)$ as per Equation~\ref{eq:RF}.
    \item Compute $max(RF)$ and update $opt\_allo$ as the ad sequence corresponding to row $j$.
    \item Repeat (4) and (5) for each row in $C$ to obtain optimal $opt\_allo$.  
\end{enumerate}\vspace{1mm}
\caption{\small Brute-force ACAD implementation.}\label{bf_acad} 
\end{algorithm}
\normalsize
\begin{figure}[!htbp]
\includegraphics[width=0.48\linewidth]{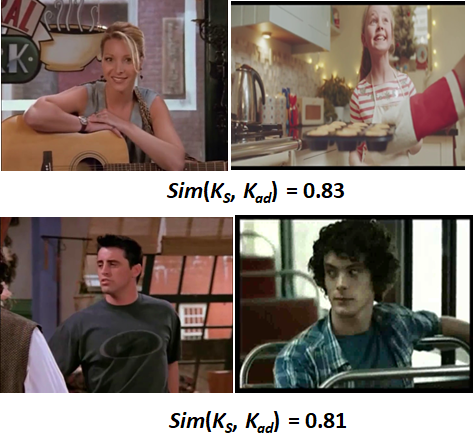}\hspace{0.02cm}\includegraphics[width=0.48\linewidth]{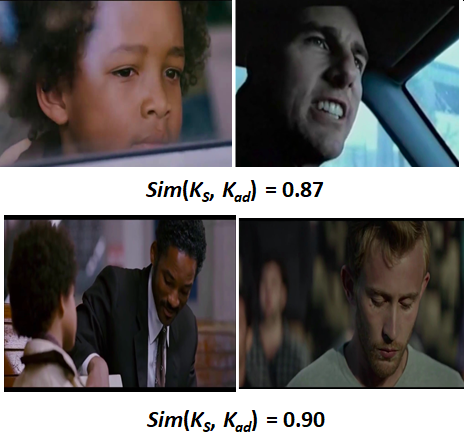}
\caption{Scene-ad keyframe pairs and similarity values for two examplars from the (left) HV and (right) LV program. }\label{fig:Sim_pairs}
\end{figure}

Algorithm~\ref{bf_acad} presents a brute-force implementation of the ACAD formulation. Input to the algorithm are scene and ad valence vectors, number of ads to be inserted, number of available ad slots and the scene-ad similarity matrix. The algorithm outputs an optimal ad-embedded VPS, denoting which ad is to be inserted at which ad-slot. The optimal sequence $opt\_allo$ of length $M$ is initialized to zero (Step $1$). Step $2$ considers all $K$ ad combinations that can be chosen from $P$ candidates. To avoid bias, only those combinations involving an equal number of HV and LV ads are deemed valid. On choosing $K$ ads, their optimal VPS placements need to be computed (Step $3$). Here, we consider all permutations of $K$ ads via matrix $L$, and retain only those placements which satisfy constraint (c). Each row of $L$ gives a matrix $Y$ satisfying all constraints of Equation \ref{eq:RF}. The reward function $RF$ is then computed, and maximum $RF$ value updated (Step $5$). Steps $4$ and $5$ are repeated for all valid ad combinations and permutations (Step $6$).

Algorithm~\ref{bf_acad} complexity is exponential in terms of the number of inventory ads and ad-slots. This is not serious in practice as in a typical VPS (\eg, \emph{Youtube}-streamed videos), the number of ads and ad slots are limited in number. Efficient ways to solve binary integer programs such as \emph{branch-and-bound} exist, and are easily solved by solvers like Gurobi~\cite{gurobi}. Another efficient way to solve Eq.(\ref{eq:RF}) is to relax the linear program allowing $y_{ij}$ to assume real values in [0,1]. Solution obtained thereof may not be optimal, and we will attempt to efficiently solve Eq.(\ref{eq:RF}) in future.

To compute the scene-ad similarity matrix, we employed the \emph{Places365 Resnet152} model~\cite{zhou2017places}. 40 uniformly sampled key-frames from each video scene and ad, denoted as $\{K_s\}$ and $\{K_{ad}\}$ are input to the Resnet and 512-D fc7 features extracted. Similarity between keyframe pairs is then computed via \emph{cosine} distance, and $rel(S_i,A_j)$ denotes the mean similarity among 40 pairs. Fig.~\ref{fig:Sim_pairs} presents exemplar video scene-ad keyframe pairs and their similarity scores. The Resnet model effectively captures similarities in scene structure (cafe-kitchen scenes for left-top pair) and face/body orientation, and enables selection of semantically similar-but-emotionally dissimilar ads to follow scenes in ACAD.

\subsection{CAVVA vs ACAD Scheduling}\label{Sec:Insert_strat}
The key difference between the CAVVA~\cite{cavva} and ACAD formulations is that while CAVVA leverages emotional inertia to embed ads that minimally perturb the user's emotional state and viewing experience, ACAD seeks to insert ads strongly incongruent to the program mood to elicit genuine ad assessments and maximal recall. To examine differences in VPSs generated by these two VIVA methods, we added four scenes to both the Friends and IPOH PVs used in Sec.~\ref{Exp:Mat} so that the new Friends and IPOH programs were  respectively about 24 and 18 minutes long. Eight ads-- four each of HV and LV were then embedded onto these PVs (with 11 ad-slots) via the CAVVA and ACAD methods. These ads were selected from distinct inventories for the two programs (see Sec.~\ref{Sec:val_us}).

\begin{figure}[!htbp]
\centerline{\includegraphics[width=0.45\linewidth,height=3.5cm]{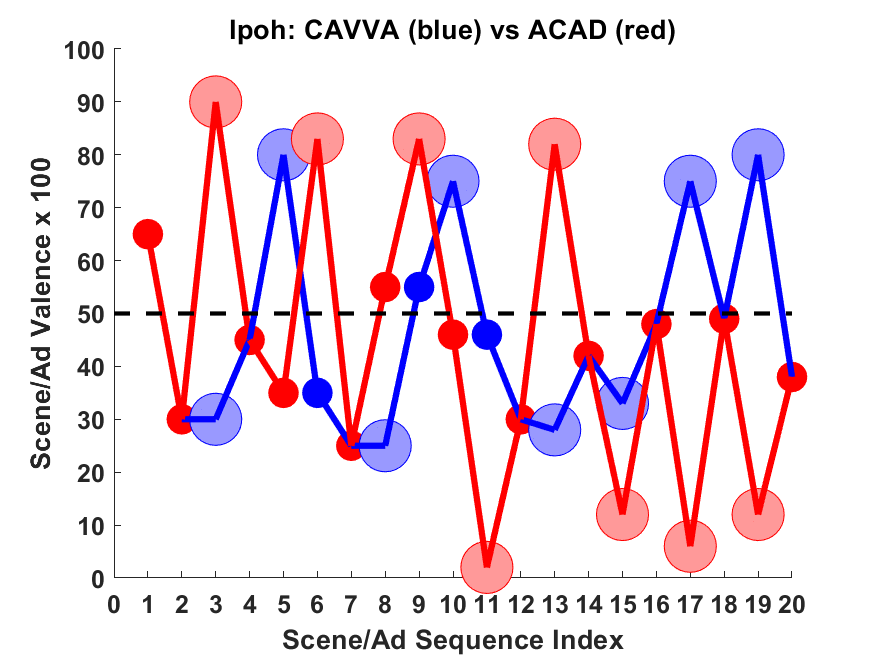}\hspace{.5mm}\includegraphics[width=0.45\linewidth,height=3.5cm]{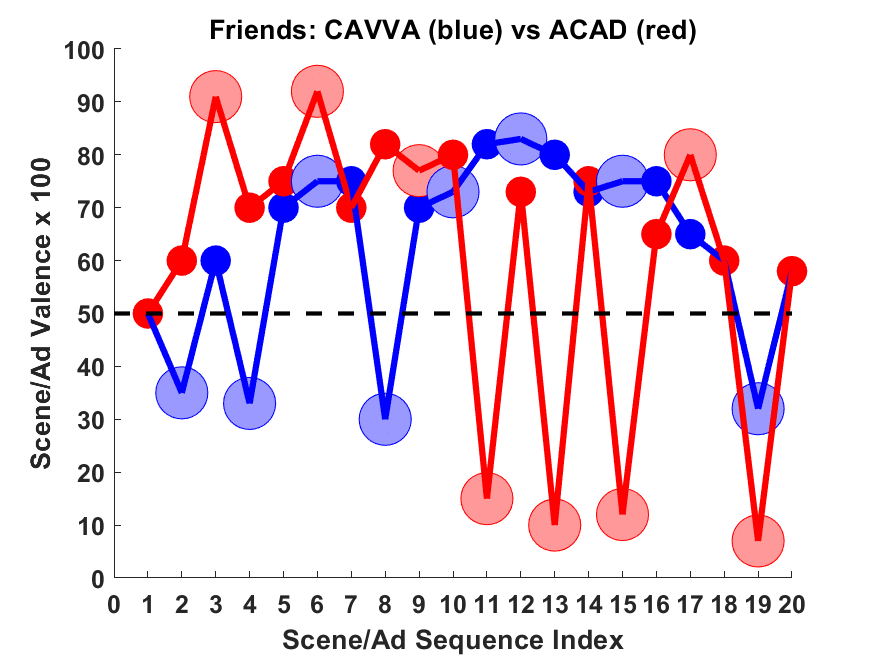}}
\caption{CAVVA (blue) and ACAD (red) VPS valence profiles for the IPOH (L) and Friends (R) programs. Eight ads are inserted within a 12 scene PV. Small solid balls denote scenes, while large hollow balls denote the inserted ads.}\label{fig:VPS_comp}
\vspace{2mm}
\centerline{\includegraphics[width=0.4\linewidth,height=4cm]{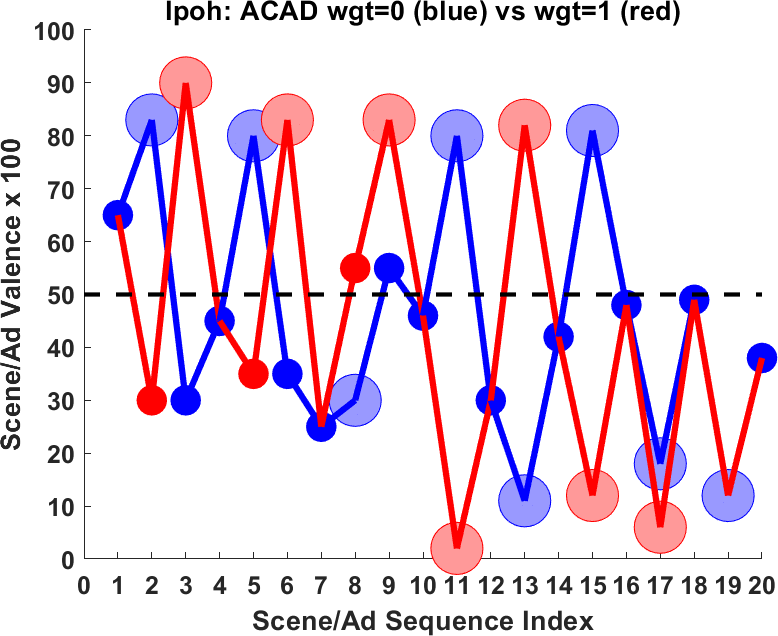}\hspace{2mm}\includegraphics[width=0.4\linewidth,height=4cm]{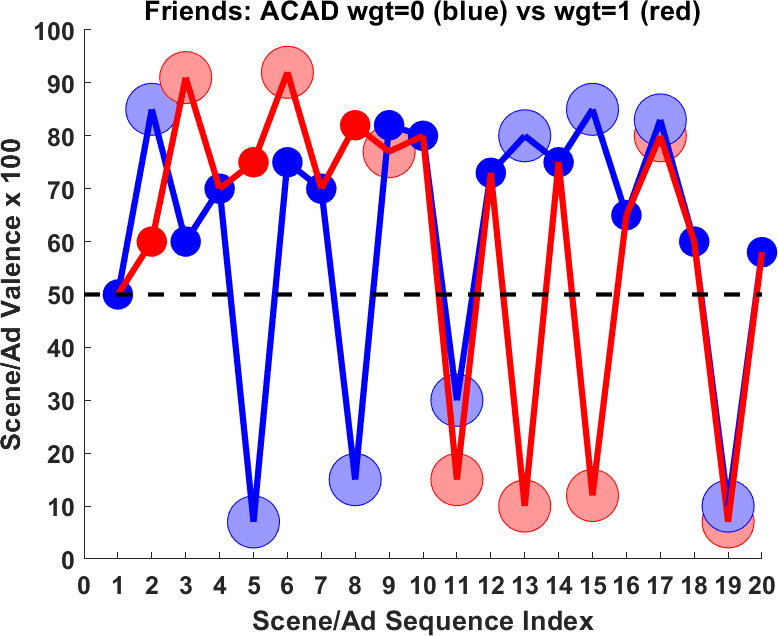}}
\caption{Val profiles for the IPOH and Friends ACAD VPSs with $\alpha/\beta=1$ (red) and $0$ (blue). Eight ads (hollow balls) were inserted within PV comprising 12 scenes (solid balls). }\label{fig:ACAD_abl} \vspace{-4mm}
\end{figure}

Fig.~\ref{fig:VPS_comp} presents val profiles (on a scale of 100) of the generated CAVVA and ACAD VPSs for the sad and happy programs. Scene/ad val scores input to the ACAD and CAVVA algorithms were computed as the average of the ratings provided by three experts. Smaller, solid balls denote program scenes while the larger, hollow balls represent the eight embedded ads. Scenes/ads having a val score of $>50$ are deemed HV, and others LV. From the generated VPSs, we make the following observations. 

\begin{itemize}
\item Clearly, CAVVA attempts to generate \emph{smooth} VPS valence profiles to maintain emotional inertia. This is especially evident for the Friends VPS (Fig.~\ref{fig:VPS_comp}(right)) where HV ads are added between scenes of similar valence so as to sustain a positive mood. For the IPOH program (Fig.~\ref{fig:VPS_comp}(left)) which contains a number of LV scenes, LV ads are added next to LV scenes, so as to enable viewers to transition to a happy emotional state through the prior scene-ad-next scene sequence. 

\item Similar to ACAD, CAVVA was constrained to include both HV and LV ads in the generated VPS, different from its natural behavior. Under such conditions, CAVVA attempts to embed HV (LV) ads in the sad (happy) programs, so that the resulting VPS valence profile is as `less spiky' as possible. Further, the emphasis on viewing experience results in LV ads being embedded earlier rather than later in the VPS. This behavior adversely impacts ad val assessment and recall as discussed in Sec.~\ref{Sec:val_us}.

\item Contrastingly, ACAD favors the generation of VPSs with \emph{spiky} val profiles consistent with our hypothesis that ads emotionally different to preceding scenes are evaluated and recalled better. The ACAD Friends and IPOH VPSs exhibit `bumpy' val profiles. ACAD formulation also forces LV ads to be embedded at the VPS tail. 
\end{itemize}

Recall that the reward function $RF$ specified by Eq.\ref{eq:RF} comprises trade-off parameters $\alpha$ and $\beta$. While higher $\alpha$ enforces the embedding of LV ads towards the VPS tail, $\beta$ emphasizes on an embedding scheme where program scenes are followed by emotionally dissimilar-but-semantically similar ads. To examine the impact of this trade-off, we generated ACAD VPSs with $\alpha = \beta = 0.5$ and $\alpha = 0, \beta = 1$. Val profiles for the corresponding VPSs are respectively shown in red and blue in Fig.~\ref{fig:ACAD_abl}. When ACAD solely focuses on semantic-plus-emotional ad-scene matching (blue curve), LV ads are also added towards the VPS \emph{head} (Fig.~\ref{fig:ACAD_abl}~(right)). Little difference is noted between the red and blue ACAD VPSs generated for the IPOH program (Fig.~\ref{fig:ACAD_abl}(left)). Nevertheless, the red curves denoting a positioning--relevance tradeoff achieve optimal ad assessments and recall as evidenced by the following validational user study. 

\section{Validational Study \& Hypotheses}\label{Sec:val_us}
\begin{sloppypar}
To confirm if the ACAD rules indeed enable optimal ad assessments and recall, we performed a validational user study comparing the CAVVA and ACAD methods against a simplistic (`Trivial') VIVA approach. Different from CAVVA and ACAD that enforce a uniform distribution of ads across the VPS (constraint (c) in Eq.\ref{eq:RF}), the Trivial method works by placing multiple ads at the VPS \emph{head} and \emph{middle}. The CAVVA and ACAD VPSs shown in Fig.~\ref{fig:VPS_comp} were used in this study. The Trivial method randomly selects four HV plus four LV ads, following which four ads each are randomly arranged before program scenes 1 and 7 respectively. In line with the design of CAVVA and ACAD, we expected
\end{sloppypar}
\begin{itemize}
\item ACAD to achieve superior ad evaluations and recall. 
\item CAVVA to induce the optimal viewing experience.
\item Both these methods to perform better than Trivial embedding in the above respects. 
\end{itemize}   


\vspace{-2mm}
\begin{figure*}[!b]
\centerline{\includegraphics[width=0.24\linewidth,height=3.2cm]{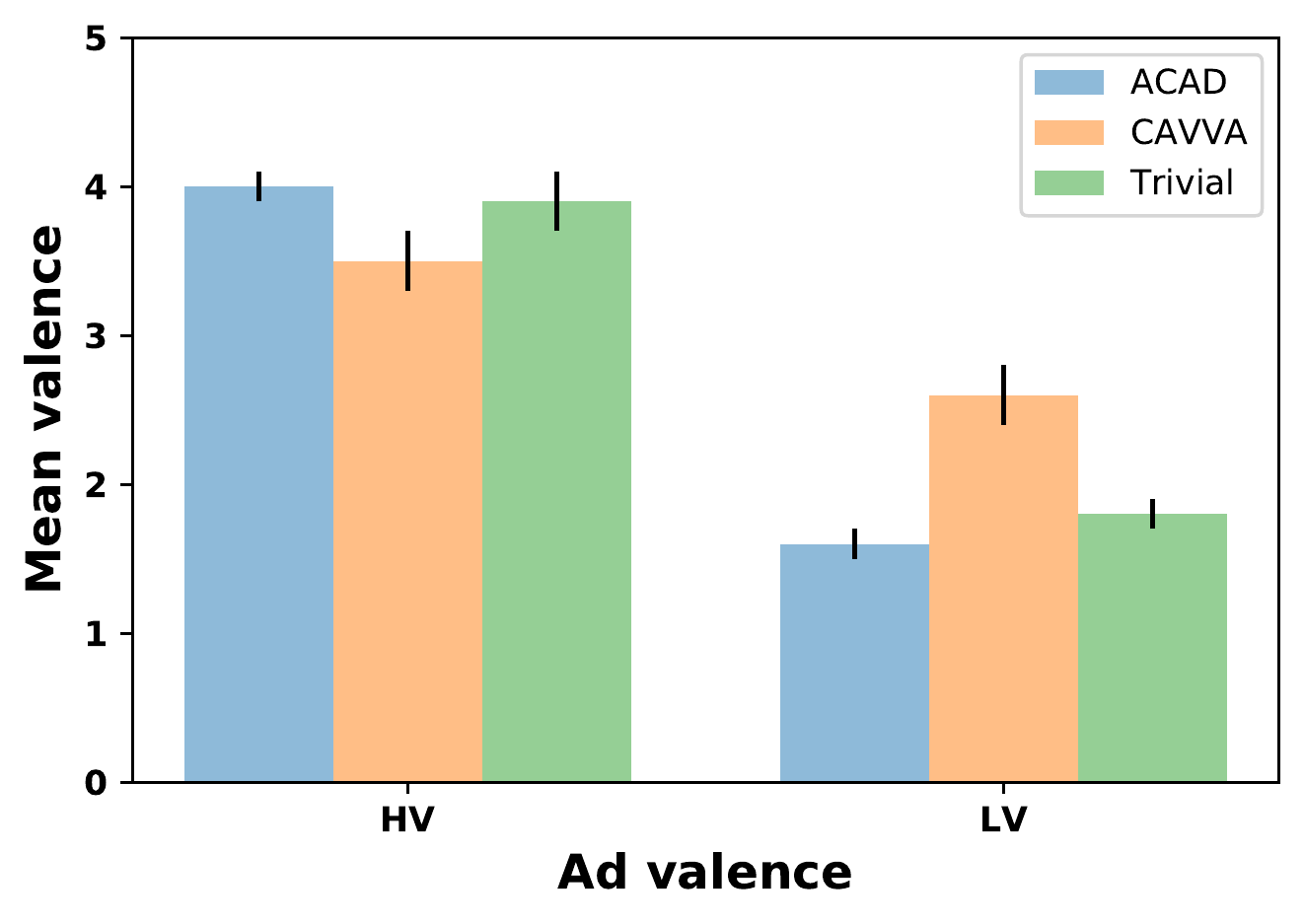}\hspace{0.02cm}\includegraphics[width=0.24\linewidth,height=3.2cm]{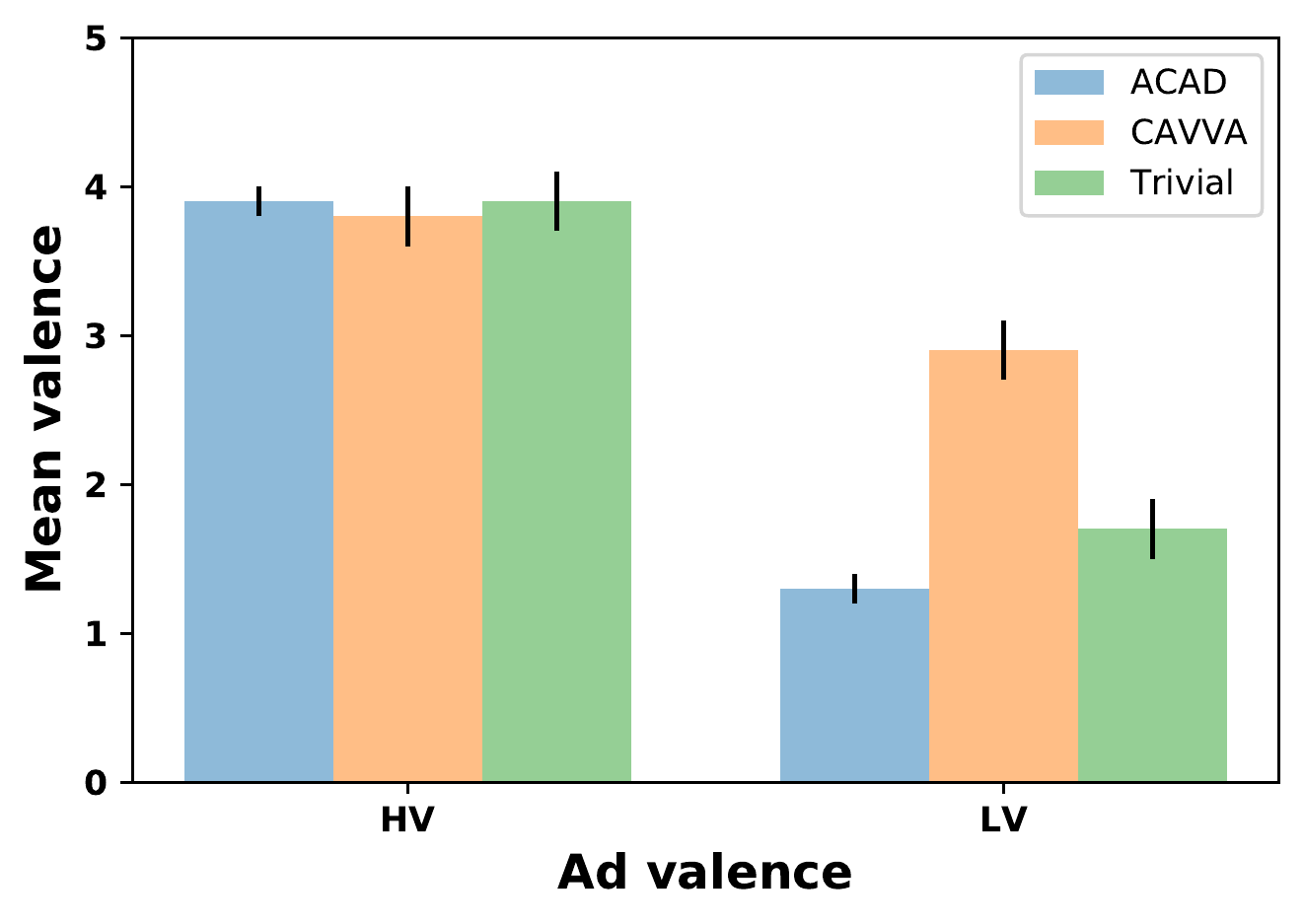}\hspace{0.02cm}\includegraphics[width=0.24\linewidth,height=3.2cm]{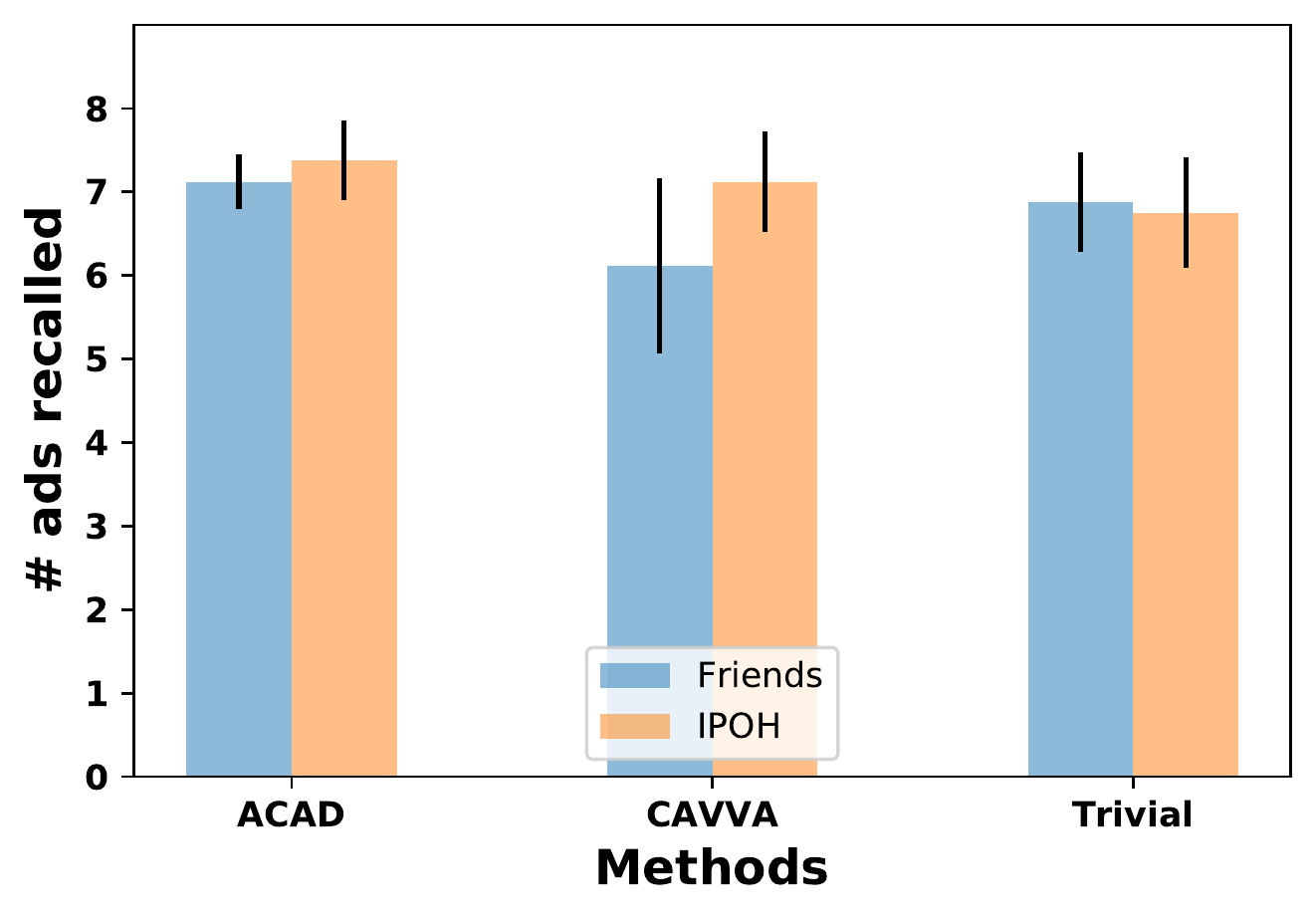}\hspace{0.02cm}\includegraphics[width=0.24\linewidth,height=3.2cm]{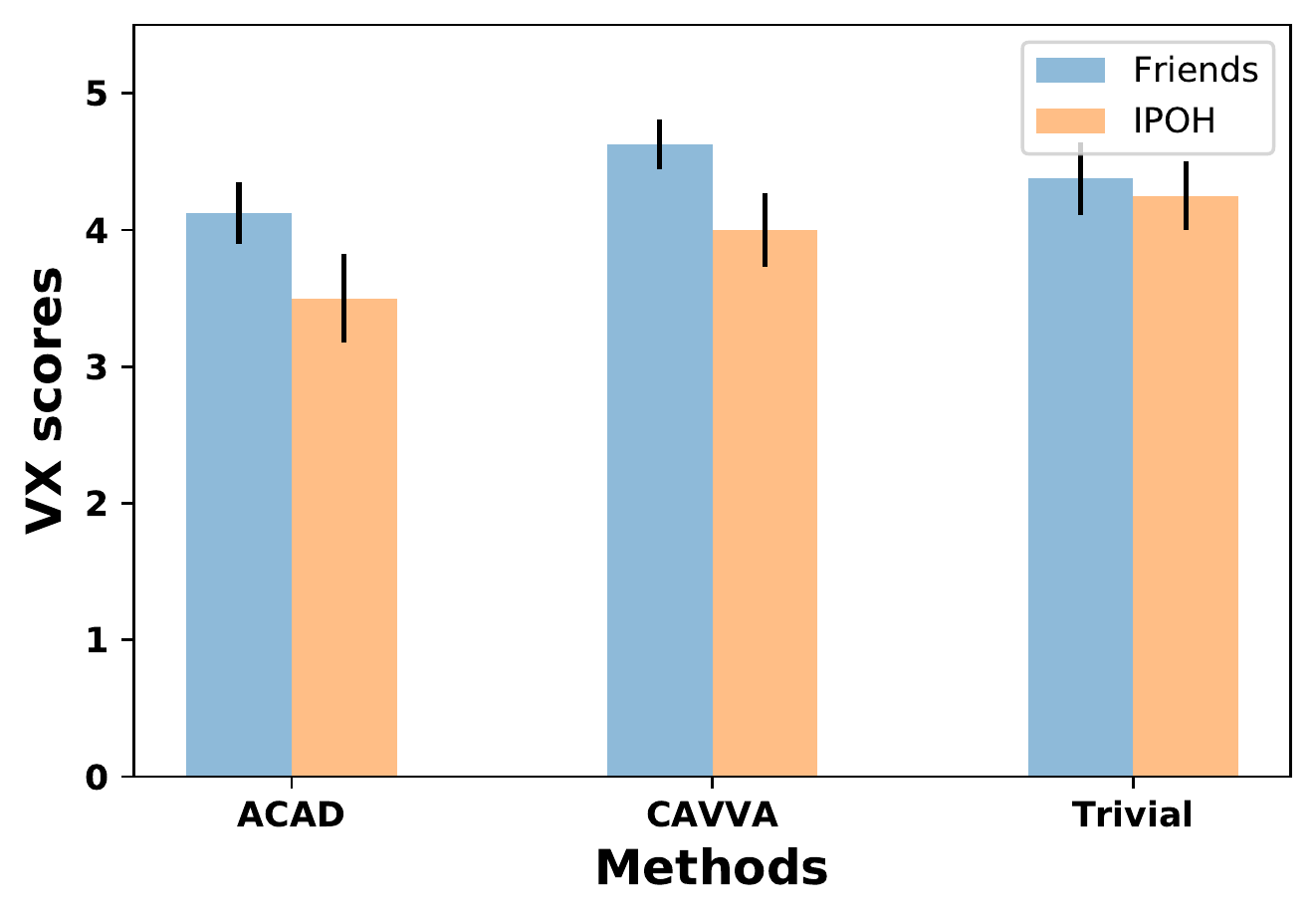}}
\caption{Validational study: (L to R) Ad val ratings for +ve and -ve VPSs, Ad recall and VX scores for the different VIVA methods.}\label{fig:VUS_comp} 
\end{figure*}

\subsection{Materials and Methods}

\textbf{Programs and Ads:} The Friends and IPOH PVs with 12 scenes each (Sec.~\ref{Sec:Insert_strat}) were used in this validational study. 48 ads (24 HAHV and 24 HALV) ads from~\cite{Shukla2020} were used. To avoid memory effects, we created distinct ad inventories for the two programs; both inventories comprised 24 ads (12 HV plus 12 LV). VPSs with eight ad embeddings were then generated for the two PVs via the ACAD, CAVVA and Trivial approaches, and presented to users. 

\noindent{\textbf{Users:}} 24 university students (14 male), naive to the purpose of the study and different from the participants in Sec.~\ref{Sec:MandM}, viewed and assessed the above VPSs (2 programs $\times$ 3 VIVA schemes = 6 VPSs). It took about one hour to cumulatively view the Friends and IPOH VPSs, and users were paid a token fee for their time and effort. 

\noindent{\textbf{Protocol:}} Each user viewed the Friends VPS generated by one VIVA method, and the IPOH VPS generated via another approach. The VPS viewing order was counterbalanced across users to ensure that (a) all VPSs had the same number of viewings, and (b) and an equal number of users viewed the happy VPS first and vice-versa. Upon viewing each VPS scene and ad, users had to rate them for val (1--5 scale); following VPS completion they were required to (i) recall the 8 viewed ads from thumbnails from the 24 inventory ads, and (ii) rate the VPS VX. Six users cumulatively viewed two instances of each of the six VPSs, resulting in 16 ad-specific, and two VPS-specific scores for recall and VX per user.

\subsection{Results}



Fig.~\ref{fig:VUS_comp} presents the user study results, with error bars denoting standard error. With respect to \emph{\textbf{ad valence evaluations}} (first two bar plots from the left), the ACAD and Trivial methods achieved more genuine assessments than CAVVA (higher val scores for HV ads, and lower scores for LV ads). For the Friends VPS, ACAD, CAVVA and Trivial HV ad embeddings received mean val ratings of 4, 3.5 and 3.9; val scores for ACAD and CAVVA differed marginally as per a two-sample \emph{t}-test $(t(14) = 1.857, p=0.093)$. LV ads inserted via ACAD evoked the lowest mean val score of 1.6; mean CAVVA and Trivial ad scores were 2.6 and 1.8 respectively. Distinct val scores were evoked for ACAD vs CAVVA LV ads ($(t(14) = -4.058, p<0.005)$), and for CAVVA vs Trivial LV ads $(t(14) = 3.237, p<0.01)$.    

For the sad IPOH VPS, HV ads embedded via all three VIVA methods evoked similar val scores (mean val score of 3.9, 3.8 and 3.9 for ACAD, CAVVA and Trivial), but distinct ratings were noted for LV ads (mean val score of 1.3, 2.9 and 1.7 for ACAD, CAVVA and Trivial). Distinct val ratings were noted for ACAD vs CAVVA ads $(t(14) = -7.439, p<0.0001)$, and slightly distinct ratings for ACAD vs Trivial $(t(14) = -1.857, p<0.096)$. Overall, while HV ad val ratings were unaffected by the VIVA approach, ACAD-embedded LV ads elicited genuine val scores than other methods.    

With respect to \emph{\textbf{ad recall}}, more ads were generally recalled for the IPOH VPS (mean recall of 7.4, 7.1 and 6.8 ads for ACAD, CAVVA and Trivial), but greater recall variations were noted for the Friends VPS (mean recall of 7.1, 6.1 and 6.9 ads for ACAD, CAVVA and Trivial). A slight difference in IPOH ad recall was noted for ACAD vs Trivial $(t(14) = 2.017, p=0.065)$, while distinct recall rates were noted for ACAD vs CAVVA $(t(14) = 2.397, p<0.05)$ methods in the Friends program. In terms of \emph{\textbf{viewing experience}}, the CAVVA-generated Friends VPS received the highest scores (mean VX = 4.6) followed by Trivial (mean VX = 4.4) and ACAD (mean VX = 4.1), while the Trivial IPOH VPS (mean VX = 4.3) received higher scores than CAVVA (mean VX = 4) and ACAD (mean VX = 3.5). Overall, (a) ACAD achieves higher ad recall at the cost of VX (due to scene-ad emotional dissimilarity), and (b) when constrained to include ads strongly opposite  to program mood, CAVVA may not necessarily achieve the optimal VX.  


\vspace{-2mm}
\section{Summary and Conclusions}
Based on the exploratory and validational user studies each involving 24 users, we show that (a) the ACAD VIVA strategy devised from user impressions  rather than axiomatic rules, is flexible and can effectively accommodate ads strongly incongruent to program mood, and (b) ACAD induces genuine ad evaluations and superior recall in users as compared to CAVVA . While ACAD rules are derived from a user-study involving two short programs and a small number of ads, these rules are nevertheless generalizable, as confirmed by the validational study involving longer programs, larger number of ads and an independent set of users. The validational study also reveals a trade-off between ad recall and viewing experience, which should guide future designs. A limitation of our work is that we ignore the arousal dimension for ad evaluation, and this will be investigated in future. Other improvements as part of future work include: (a) designing a more efficient ad-embedding strategy than a brute-force approach, and (b) evaluating ACAD in real-life situations where scene and ad ratings are computed via automated methods. 

\bibliographystyle{unsrt}  
\bibliography{references}  

\begin{thebibliography}{10}

\bibitem{cavva}
Karthik Yadati, Harish Katti, and Mohan Kankanhalli.
\newblock {CAVVA: Computational affective video-in-video advertising}.
\newblock {\em IEEE Trans. Multimedia}, 16(1):15--23, 2014.

\bibitem{Mei09}
T.~{Mei}, X.~{Hua}, and S.~{Li}.
\newblock Videosense: A contextual in-video advertising system.
\newblock {\em IEEE Transactions on Circuits and Systems for Video Technology},
  19(12):1866--1879, 2009.

\bibitem{Russell1980}
James Russell.
\newblock {A circumplex model of affect}.
\newblock {\em Journal of Personality and Social Psychology}, pages 1161--1178,
  1980.

\bibitem{Broach1995TelevisionPA}
V.~Broach, Thomas~J. Page, and R.~Wilson.
\newblock Television programming and its influence on viewers' perceptions of
  commercials: The role of program arousal and pleasantness.
\newblock {\em Journal of Advertising}, 24(4):45--54, 1995.

\bibitem{Holbrook1984}
Morris~B Holbrook and John~O Shaughnessy.
\newblock {The role of emotlon in advertising}.
\newblock {\em Psychology {\&} Marketing}, 1(2):45--64, 1984.

\bibitem{Holbrook1987}
Morris~B Holbrook and Rajeev Batra.
\newblock {Assessing the Role of Emotions as Mediators of Consumer Responses to
  Advertising}.
\newblock {\em Journal of Consumer Research}, 14(3):404--420, 1987.

\bibitem{Pham2013}
Michel~Tuan Pham, Maggie Geuens, and Patrick~De Pelsmacker.
\newblock The influence of ad-evoked feelings on brand evaluations: Empirical
  generalizations from consumer responses to more than 1000 {TV} commercials.
\newblock {\em Int'l Journal of Research in Marketing}, 30(4):383 -- 394, 2013.

\bibitem{Shukla2020}
A.~{Shukla}, S.~S. {Gullapuram}, H.~{Katti}, M.~{Kankanhalli}, S.~{Winkler},
  and R.~{Subramanian}.
\newblock Recognition of advertisement emotions with application to
  computational advertising.
\newblock {\em IEEE Transactions on Affective Computing}, 2020.

\bibitem{Shukla18}
Abhinav Shukla, Harish Katti, Mohan Kankanhalli, and Ramanathan Subramanian.
\newblock Looking beyond a clever narrative: Visual context and attention are
  primary drivers of affect in video advertisements.
\newblock In {\em ACM International Conference on Multimodal Interaction}, page
  210–219, 2018.

\bibitem{FRIESTAD19931}
Marian Friestad and Esther Thorson.
\newblock Remembering ads: The effects of encoding strategies, retrieval cues,
  and emotional response.
\newblock {\em Journal of Consumer Psychology}, 2(1):1 -- 23, 1993.

\bibitem{Subramanian14}
Ramanathan Subramanian, Divya Shankar, Nicu Sebe, and David Melcher.
\newblock {Emotion modulates eye movement patterns and subsequent memory for
  the gist and details of movie scenes}.
\newblock {\em Journal of Vision}, 14(3):31--31, 03 2014.

\bibitem{Kamins91}
Michael Kamins, Lawrence Marks, and Deborah Skinner.
\newblock Television commercial evaluation in the context of program induced
  mood: Congruency versus consistency effects.
\newblock {\em Journal of Advertising}, 20(2):1--14, 2013.

\bibitem{Meyers1993}
Joan Meyers-Levy and Brian Sternthal.
\newblock A two-factor explanation of assimilation and contrast effects.
\newblock {\em Journal of Marketing Research}, 30(3):359--368, 1993.

\bibitem{Aylesworth98}
Andrew~B. Aylesworth and Scott~B. MacKenzie.
\newblock Context is key: The effect of program-induced mood on thoughts about
  the ad.
\newblock {\em Journal of Advertising}, 27(2):17--31, 1998.

\bibitem{Feltham1994}
Tammi Feltham and Stephen Arnold.
\newblock Program involvement and ad/program consistency as moderators of
  program context effects.
\newblock {\em Journal of Consumer Psychology}, 3:51--77, 12 1994.

\bibitem{Furnham13}
Adrian Furnham and Mei Goh.
\newblock Effects of program-advertisement congruity and advertisement
  emotional appeal on memory for health and safety advertisements.
\newblock {\em Journal of Applied Social Psychology}, 44, 12 2013.

\bibitem{Moorman07}
Marjolein Moorman, Peter Neijens, and Edith Smit.
\newblock The effects of program involvement on commercial exposure and recall
  in a naturalistic setting.
\newblock {\em Journal of Advertising - J ADVERTISING}, 36:125--141, 04 2007.

\bibitem{shapiro2013understanding}
Stewart Shapiro, J.~Deborah Macinnis, and Whan~C. Park.
\newblock Understanding program induced mood effects: Decoupling arousal from
  valence.
\newblock {\em Journal of Advertising}, 2013.

\bibitem{Yun2020}
Joseph~T. Yun, Claire~M. Segijn, Stewart Pearson, Edward~C. Malthouse,
  Joseph~A. Konstan, and Venkatesh Shankar.
\newblock Challenges and future directions of computational advertising
  measurement systems.
\newblock {\em Journal of Advertising}, 49(4):446--458, 2020.

\bibitem{Kensinger}
Elizabeth~A. Kensinger.
\newblock Remembering the details: Effects of emotion.
\newblock {\em Emotion Review}, 1(2):99--113, 2009.

\bibitem{Rimmele11}
Ulrike Rimmele.
\newblock Emotion enhances the subjective feeling of remembering, despite lower
  accuracy for contextual details.
\newblock {\em Journal of Vision}, 11(3), 2011.

\bibitem{gurobi}
LLC Gurobi~Optimization.
\newblock Gurobi optimizer reference manual, 2021.

\bibitem{zhou2017places}
Bolei Zhou, Agata Lapedriza, Aditya Khosla, Aude Oliva, and Antonio Torralba.
\newblock Places: A 10 million image database for scene recognition.
\newblock {\em IEEE Transactions on Pattern Analysis and Machine Intelligence},
  2017.

\end{thebibliography}

\end{document}